\documentclass{elsarticle}

\usepackage{lineno,hyperref}
\usepackage{pdfpages}
\usepackage{graphicx}
\usepackage{lscape}
\usepackage{supertabular}
\usepackage{graphicx, subfigure}
\usepackage{rotating}
\usepackage{enumerate}
\usepackage{color}
\usepackage{xcolor}
\usepackage{colortbl}
\usepackage{rotating}
\def\0{\mbox{\bf{0}}}
\def\bs{\mathbf{s}}\def\be{\mathbf{e}}

\def\diag{\mbox{diag}}

\usepackage[toc,page]{appendix}

\def \be{\begin{equation}}
\def \ee{\end{equation}}
\def \ber{\begin{eqnarray}}
\def \eer{\end{eqnarray}}
\def \berr{\begin{eqnarray*}}
\def \eerr{\end{eqnarray*}}
\def \bqmatrix{\begin{bmatrix}}
\def \eqmatrix{\end{bmatrix}}

\newtheorem{proposition}{Proposition}

\def \be{\begin{equation}}
\def \ee{\end{equation}}
\def \ber{\begin{eqnarray}}
\def \eer{\end{eqnarray}}
\def \berr{\begin{eqnarray*}}
\def \eerr{\end{eqnarray*}}
\def \bamatrix{\begin{pmatrix}}
\def \eamatrix{\end{pmatrix}}
\def \bqmatrix{\begin{bmatrix}}
\def \eqmatrix{\end{bmatrix}}

\def \argmin{{\rm argmin }}

\def \argmin{{\rm argmin }}
\def \argmax{{\rm argmax }}

\def \bs{\boldsymbol}

\usepackage[latin1]{inputenc}
\usepackage[T1]{fontenc}
\usepackage{amsmath}
\usepackage{amsfonts}
\usepackage{amssymb}
\usepackage{graphicx}
\usepackage{lscape}
\usepackage{supertabular}
\usepackage{graphicx, subfigure}

\makeatletter
\def\ps@pprintTitle{%
 \let\@oddhead\@empty
 \let\@evenhead\@empty
 \def\@oddfoot{}%
 \let\@evenfoot\@oddfoot}
\makeatother

\begin{document}

\linespread{2}

\begin{frontmatter}

\title{{Joint estimation of conditional quantiles in multivariate linear regression models. An application to financial distress.}}

\author[LP]{Lea Petrella}
\ead{lea.petrella@uniroma1.it}

\author[IC,LP]{Valentina Raponi \corref{corr}}
\ead{v.raponi13@imperial.ac.uk}
\cortext[cor]{Corrisponding Author: Valentina Raponi, Imperial College Business School, South Kensigton Campus}

\address[LP]{MEMOTEF, Sapienza, Universit\`a di Roma, Rome, Italy}

\address[IC]{Imperial College Business School, Imperial Colege London, London, UK}

%

\begin{abstract}
{This paper proposes a maximum-likelihood approach {to jointly estimate marginal conditional quantiles of multivariate response variables in a linear regression framework.}
 We consider a {slight reparameterization} of the Multivariate Asymmetric Laplace distribution proposed by Kotz et al (2001) {and exploit its location-scale mixture representation} to implement a new EM algorithm for estimating model parameters. The idea is to {extend} the link between the Asymmetric Laplace distribution {and the well-known univariate quantile regression model} to {a multivariate context, i.e. when a multivariate dependent variable is concerned}. The approach accounts for association among multiple responses and study how the relationship between responses and explanatory variables can vary across different quantiles of the marginal conditional distribution of the responses. A penalized version of the EM algorithm is also presented to {tackle} the problem of variable selection. The validity of our approach is analyzed in a simulation study, where we also  provide evidence on the efficiency gain of the proposed method compared to estimation obtained by separate univariate quantile regressions. A real data application is finally proposed to study the main determinants of financial distress in a sample of Italian firms.}
\end{abstract}


\end{frontmatter}


\section{Introduction}

\noindent Quantile regression has  become a widely used technique in many empirical applications, since the seminal work of Koenker and Basset (1978). {It provides a way to
model the conditional quantiles of a response variable with respect to a set of covariates in
order to have a more complete picture of the entire conditional distribution than the ordinary
least squares regression. This approach is quite suitable to be used in all the situations where
specific features, like skewness, fat-tails, outliers, truncation, censoring and heteroscedasticity
arise. In fact,} unlike standard linear regression models, which only consider the conditional mean of a response variable, quantile regression allows one to assume that the relationship between the response and explanatory variables can vary across the conditional distribution of the dependent variable.


\noindent Many {univariate} quantile regression methods are now well consolidated in the literature and have been implemented in a wide range of different fields, like medicine (see e.g., Cole and Green (1992) and Royston and Altman (1994), Marino et al. (2016)), survival analysis (Koenker and Geling (2001)), financial and economic research (Bassett and Chen (2001), Hendricks and Koenker (1992), Petrella et al.(2017)), and environmental modelling (see Pandey and Nguyen (1999) and Hendricks and Koenker (1992) for a discussion). Koenker (2005) provides an overview of the most used quantile regression techniques in a classical setting. 
In longitudinal studies, quantile regression models with random effects are also analysed, in order to account for the dependence between serial observations on the same subject. See, e.g Geraci and Bottai (2007), Koenker (2004), Koenker (2017), Marino and Farcomeni (2015) for references. 
 Bayesian versions of quantile regression have also been extensively proposed (see Yu and Moyeed (2001), Kottas and Gelfand (2001), Kottas and Krnjajic (2009), Bernardi et al. (2015)). 

\noindent It is well known in the literature that the univariate quantile regression approach has a direct link with the Asymmetric Laplace (AL) distribution. 
{In fact, while frequentist quantile regression framework relies on the minimization of the asymmetric loss function introduced by Koenker and Basset (1978), the Bayesian approach introduces the Asymmetric Laplace (AL) distribution  as inferential tool to estimate model parameters (see the seminal work Yu and Moyeed (2001)). {The two approaches are both justified by the well-established relationship between the   loss function and the AL density. That is, the loss function minimization problem is equivalent (in terms of parameter estimates) to the maximization of the likelihood  associated with the AL density (see e.g Kozumi and Kobayashi (2011)). Therefore, the AL distribution could offer a convenient device to implement a likelihood-based inferential approach when dealing with quantile regression analysis.}}

\noindent{Still in the context of univariate regression framework, part of the literature is concerned with estimation of \textit{multiple} quantiles (say ${\cal Q}_Y(\tau_j)$, j=1,.., p) of a given response variable $Y \in {\cal R}$. In this case, joint estimation of these $p$ quantiles provides a gain in efficiency compared to traditional sequential estimation of multiple quantiles (see Koenker and Basset, 1978). A recent extension to a longitudinal setting has been proposed by Cho et al. (2017).  }

\noindent {When \textit{multivariate} response variables are concerned, the existing literature on quantile regression is  less extensive.  }
The \textit{multivariate} quantile problem  has the goal to estimate  the multivariate quantile, ${\cal Q}_{\bs Y}(\tau)$, of a multivariate response variable $\bs Y \in {\cal R}^p$, with $p>1$ and where the index $\tau$ is a scalar. In this case, the main challenge concerns the definition of a \textit{multivariate quantile}, given that there is no natural ordering in a $p$-dimensional space (Chakraborty (2003), Hallin et al. (2010), Kong and Mizera (2012)). Other attempts and extensions can be also found in   Kong et al. (2015), Bernardi et al. (2018), Paindaveine and Siman (2012), Bocek and Siman (2017) and  Alf\'o et al. (2016).

\noindent {The main goal of the present paper is to extend the univariate linear quantile regression methodology to a multivariate context. In particular we want to genereralize the  inferential approach based on the AL distribution to a multivariate framework, by using the Multivariate Asymmetric Laplace (MAL) distribution  introduced by Kotz et al. (2001). In this way we are not concerned to define a multivariate quantile.  Instead, we are conducting a simultaneous inference on the marginal conditional quantiles of a multivariate response variable,  taking also into account for the possible correlation among marginals.}

{
\noindent This need could arise in many situations where different responses might have similar distributions and/or might be affected by the same set of covariates at different parts of their distributions other than the mean. Hence, by jointly modeling them we can borrow information across
responses and conduct joint inference at the marginal quantiles level, rather then defining a central point for their distributions.} 
 
\noindent{Similar attempts in the literature have been proposed by Jun and Pinkse (2009), who introduce a  seemingly unrelated quantile regression approach which entails a nonparametric estimation of a set of moment conditions for the conditional quantiles of interest.  Waldmann and Kneib (2015) propose a bivariate quantile regression using a Bayesian approach and show how to estimate the conditional correlations between the two response variables. Their work, however, has not been generalized to the multivariate case. }

 \noindent {Our paper offers a likelihood-based approach for modeling and estimating conditional marginal quantiles jointly, by using the MAL distribution as working likelihood in a linear regression framework.}
 {Specifically, we  propose a slight reparametrization of the MAL distribution, subject to some specific constraints, which allows us to estimate regression coefficients via maximum likelihood (ML), accounting for the possible association among the responses. The inferential problem is solved by developing a suitable Expectation-Maximization (EM) algorithm, which exploits the mixture representation of the MAL distribution (see also Arslan, 2010).}

\noindent {Using simulation exercises, we assess the validity and the robustness of our approach, by considering different model distributional settings. We find that {the estimation of the regression coefficients is not highly affected by the MAL distributional assumption.}} 
\noindent {Moreover, the estimation efficiency of our multivariate approach is higher than the one obtained by running separate single quantile regression models on the marginals when estimating model parameters. That is, taking into account for the potential association among the response variables can significantly reduce the root mean square error of the estimated coefficients, hence improving the precision of the estimates.}  

\noindent {When dealing with multivariate regression, the high dimensionality setting is an intrinsic part of the model builiding problem. In order to gain in parsimony and to conduct a variable selection procedure, we consider the penalized Least Absolute Shrinkage and Selecting Operator (LASSO) approach proposed by Tibshirani (1996). In particular,  we propose a penalized version of the EM algorithm (PEM) accounting for an $l_1$ penalty term. 
We evaluate the estimation performance of the proposed approach through a simulation exercise, where we compute the bias and the root mean square error of the estimated parameters at different quantile levels.}
\noindent {The relevance of our approach is also shown empirically, contributing to the increasingly widespread literature that uses quantiles as measures of risk. }
\noindent {In the recent years, due the financial and economic crisis, a particular attention has been devoted to measuring and quantifying the level of financial risk and financial distress within a firm or investment portfolios. In this respect, many risk measures developed in literature are based on quantile values,  like for example the Value at Risk (see Jorion (2007) and McNeil at al. (2005)).  Moreover,  quantile regression methods  turn out to be very helpful to  quantify either the magnitude and the causes of riskness (see for example {Engle and Manganelli (2004), Wong and Ting (2016)}). In this paper, we implement the proposed quantile regression approach to investigate the main determinants of financial distress on a sample of 2,020 Italian firms.} In particular, we use the definition of financial distress adopted in Bastos and Pindado (2013), which classify a firm as financial distressed if its  earnings before interest and taxes depreciation and amortization (EBITDA) is under the first quartile of the sample or if its leverage is above the third quartile of the sample. Hence, starting from this definition, we apply our methodology  to analyze the relationships between financial distress and firms'
characteristics and evaluate how it may vary when considering different (more extreme) quantiles of the distribution of leverage and EBITDA. In this way we are able to assess not only what are the main determinants for a firm's risk of financial distress, but also how these factors matter as more serious levels of distress are considered.  

\noindent {The rest of the paper is organized as follows. In Section \ref{Preliminary}, we introduce the main notation and briefly revise the univariate quantile regression model. Section \ref{MultQuantRegressionMAL} introduces the joint quantile regression framework, while Section \ref{EM MLE}     proposes the EM-based Maximum Likelihood approach and the related Penalized EM (PEM) algorithm to estimate model parameters. In Section \ref{Simulation} we provide  simulation results, while the empirical application is presented in Section \ref{Empirical application}. Section \ref{Conclusions} summarizes our conclusions. }

\section{{Preliminaries on univariate quantile regression and the AL distribution}} \label{Preliminary}

\noindent {To better explain the link between the MAL distribution and the joint quantile regression, we briefly revise the univariate quantile regression model and its direct connection with the AL density. As argued in Yu and Moyeed (2001) and Kozumi and Kobayashi (2011), we say that a random variable $Y$ is distributed as an AL with location parameter $\mu$, scale parameter $\delta >0$ and skewness parameter $\tau \in (0,1)$, i.e. $AL (\mu, \delta, \tau)$, if its probability density function has the following representation
\ber \label{ALDquant}
f_{AL}(y; \mu, \delta, \tau) = \frac{\tau (1-\tau)}{\delta} \exp \left\{ - \rho_{\tau}\left( \frac{y- \mu}{\delta} \right) \right\}
\eer
\noindent where $\rho_{\tau}(\cdot)$ denotes the so called loss (or check) function defined by 
$\rho_{\tau}(x) = x \left( \tau - I\{x<0 \} \right)$, 
\noindent with $I \{ \cdot \}$ being the indicator function and where the quantity $ \rho_{\tau}\left( \frac{y- \mu}{\delta} \right)$ follows an exponential distribution with rate parameter equal to $1/\delta$.} {Kotz et al.  (2001) show that the AL distribution in (\ref{ALDquant}) admits a Gaussian-mixture representation. In particular, if $Y \sim AL (\mu, \delta, \tau)$, then $Y$ can be also written as
\ber
Y = \mu + \xi_{\tau} U + \theta_{\tau} \sqrt{\delta U} Z 
\eer
\noindent where $U$ follows an Exponential distribution with rate parameter $1/\delta$ and $Z$ is a standard Normal random variable. Moreover, in order to guarantee that the parameter $\mu$ coincides with the quantile of $Y$ at a chosen level $\tau$, the following conditions on $\xi_{\tau}$ and $\theta_{\tau}$ must be satisfied
\ber
\xi_{\tau}= \frac{1 - 2 \tau}{\tau (1 - \tau)}, \qquad and  \qquad \theta_{\tau}= \frac{2}{\tau (1- \tau).}
\eer
\noindent Now, let $y_i$, $i=1,2,...,n$ be a response variable of interest and let $\bs x_i$ be a $k \times 1$ vector of covariates associated with the $i$-th observation. Let ${\cal Q}_{y_i}\left( \tau| \bs x_i \right)$ denote the quantile regression function of $y_i$ given $\bs x_i$ at a given level $\tau \in (0,1)$, and assume that the relationship between ${\cal Q}_{y_i}\left( \tau| \bs x_i \right)$ and $\bs x_i$ can be modeled as 
\ber \label{univariate quantile}
{\cal Q}_{y_i}\left( \tau| \bs x_i \right) = \bs x_i' \bs{\beta}_{\tau}
\eer where $ \bs{\beta}_{\tau}$ is a $k \times 1$ vector of regression coefficients. Notice that the relationship in (\ref{univariate quantile}) implies the following linear quantile regression model
\ber \label{univariate quantile regression}
y_i = \bs x_i' \bs{\beta}_{\tau} + \epsilon_i, \qquad i=1,2,...,n 
\eer 
where the error term $\epsilon_i$ is such that its distribution is restricted to have the $\tau$-quantile equal to zero. If the distribution of the error term is left unspecified, then the parameter estimation proceeds by minimizing the following objective function
\ber \label{univariate loss}
\bs{\hat \beta}_{\tau} = \underset{\bs{\beta} \in {\cal R}^k}{\argmin} \sum_{i=1}^n \rho_{\tau}\left( y_i - \bs x_i' \bs{\beta}_{\tau}  \right).
\eer  
As the loss function $\rho_{\tau}(\cdot)$ is not differentiable at zero, explicit solutions for $\bs{}\hat{\beta}_{\tau}$ cannot be derived and direct optimization is typically applied. As shown in Yu and Moyeed (2001) the AL distribution provides a direct connection  between the minimization problem in (\ref{univariate loss}) and maximum likelihood (ML) estimation.  In fact, if we {use the $AL$ density as likelihood tool in (\ref{univariate quantile regression}), we have}
%
\ber \label{univariate likelihood}
{\cal L}\left( \bs{\beta}_{\tau}, \delta | \bs{y} \right) = \frac{\tau^n (1 - \tau)^n}{\delta^n} \exp \left\{- \sum_{i=1}^n \rho_{\tau}\left( \frac{y_i - \bs x_i'\bs{\beta}_{\tau}}{\delta} \right)  \right\}.
\eer
for a given $\tau$, and with  $\delta>0$. It is easy to verify that the minimization of the objective function in (\ref{univariate loss}) with respect to the parameter $\bs \beta_{\tau}$ is equivalent to the maximization of the likelihood in (\ref{univariate likelihood}). Therefore, the AL distribution offers a valid tool to set up the quantile regression model in a likelihood framework.}

\noindent {In the next section we extend such link between the AL and the quantile regression to a multivariate framework.}

\section {Joint Quantile Regression and the MAL distribution}\label{MultQuantRegressionMAL}
\noindent {Extending the results of the previous section,  we now show how to use the MAL distribution (Kotz et al. (2001)) for jointly modeling marginal conditional quantiles of a multivariate response variable.} 


\noindent {Let  $\bs Y_i=[Y_{i1}, Y_{i_2},..., Y_{ip}]'$ be a $p$-variate response variable for each individual $i=1,2,...,n$ and assume that the $\tau_j$-quantile of each of the $j$-th  component of $\bs Y_i$ can be modeled as a function of some $k$ independent variables, for $j=1,2,...,p$.   Let $\bs X_i$ be a $k \times 1$ vector of regressors for the $i$-th observation and let  $\bs\beta_{\tau}= \left[ \bs \beta_{\tau_1} , \bs \beta_{\tau_2},..., \bs \beta_{\tau_j},...,\bs \beta_{\tau_p} \right]'$ be a $p \times k $ matrix of unknown parameters, with $\bs \beta_{\tau_j}= [\beta_{1,\tau_j}, \beta_{2,\tau_j},..., \beta_{k, \tau_j}]$. Then, assume that the relationship between $\bs Y_i$ and $\bs X_i$ can be modeled as follows:
\ber \label{margQm}
\bqmatrix {\cal Q}_{Y_{i1}}(\tau_1|\bs X_i) \\ {\cal Q}_{Y_{i2}}(\tau_2|\bs X_i) \\ \vdots \\{\cal Q}_{Y_{ip}}(\tau_p|\bs X_i)  \eqmatrix &=&\bs  \beta_{\tau} \bs X_i,   \quad \quad i=1,2,...n
\eer
\noindent where ${\cal Q}_{Y_{ij}}(\tau_j|\bs X_i)$ denotes the $\tau_j$-level quantile regression function of $Y_{ij}$ given $\bs X_i$.  
Our objective is to provide  joint estimation of  the $p$ marginal conditional
quantiles of $\bs Y_i \in {\cal R}^p$. The representation in  (\ref{margQm}) implies the following multivariate linear regression model:
\ber\label{multRegr}
\bs Y_i = \bs \beta_{\tau} \bs X_i + \bs \epsilon_i \quad \quad i=1,2,...n
\eer
\noindent where $\bs \epsilon_i$ denotes a $p \times 1$ vector of error terms with {univariate} component-wise quantiles (at fixed levels $\tau_1,.., \tau_p$, respectively) equal to zero.}  \\
{For the regression model  in (\ref{multRegr}),  consider now  the following 
$\mbox{MAL}_p \left(\bs \beta_{\tau} \bs X_i,  \bs D \bs {\tilde\xi}, \, \bs D \bs {\tilde\Sigma} \bs D  \right)$ distribution with density function (see Kotz et al., 2001)
\ber \label{MALdensityCons}
f_Y(\bs y_i | \bs \beta_{\tau} \bs X_i , \bs D\bs {\tilde\xi}, \bs D\bs {\tilde\Sigma} \bs D) =  \frac{ 2 
\exp{\left\{(\bs y_i- \bs \beta_{\tau} \bs X_i)' \bs D^{-1} \bs{\tilde \Sigma}^{-1}\bs {\tilde\xi} \right\}}}
{(2\pi)^{p/2} |\bs D \bs {\tilde \Sigma} \bs D|^{1/2}   } \left( \frac{\tilde m}{2+\tilde d}\right)^{\nu/2}K_{\nu}\left(  \sqrt{(2+\tilde d)\tilde m} \right)
\eer
where $\beta_{\tau} \bs X_i $ is the location parameter vector,   $\bs D \bs {\tilde\xi} \in {\cal R}^p$ is the scale (or skew) parameter, with $\bs D= \diag [\delta_1, \delta_2,..., \delta_p]$, $\delta_j>0$ and $ \bs {\tilde\xi}= [\tilde \xi_1, \tilde \xi_2,...,\tilde  \xi_p]'$, having generic element  $\tilde \xi_j= \frac{1- 2 \tau_j}{\tau_j(1 - \tau_j)}$. $\bs {\tilde \Sigma}$ is a $p \times p$ positive definite matrix such that $\bs {\tilde \Sigma} = \bs{\tilde \Lambda} \bs \Psi \bs{\tilde \Lambda}$, with $\bs \Psi $ being a correlation matrix and $\bs{\tilde \Lambda}= \diag[\tilde \sigma_1, \tilde \sigma_1,..., \tilde \sigma_p]$, with $\tilde \sigma_j^2= \frac{2}{\tau_j (1 - \tau_j)}$, $j=1,..., p$. Moreover, $\tilde m= (\bs y- \bs   \beta_{\tau} \bs X_i)' (\bs D \bs {\tilde \Sigma }\bs D)^{-1}(\bs y-  \bs \beta_{\tau} \bs X_i)$, $\tilde d=\bs {\tilde\xi}' \bs{ \tilde \Sigma} \bs{\tilde \xi}$, and  $K_{\nu}(\cdot)$ denotes the modified Bessel function of the third kind with index parameter $\nu= (2-p)/2$. 
}
\noindent {Using  (\ref{multRegr}) and (\ref{MALdensityCons}), and following Kotz et al. (2001),  the $\mbox{MAL}_p \left(\bs \beta_{\tau} \bs X_i,  \bs D \bs {\tilde\xi}, \, \bs D \bs {\tilde\Sigma} \bs D  \right)$ can be written as a location-scale mixture, having the following representation
\ber \label{mixtureALD}
\bs Y_i=  \bs \beta_{\tau} \bs X_i   + \bs D \bs {\tilde \xi} W + \sqrt{W}  \bs D \bs {\tilde\Sigma}^{1/2}\bs Z
\eer  }

\noindent {where $\bs Z \sim {\cal N}_p(\bs 0_p, \bs I_p)$ denotes a $p$-variate standard Normal distribution and  $W \sim \mbox{Exp}(1)$ has a standard Exponential distribution, with  $\bs Z$ being independent of $W$. }


\noindent{It is worth noticing that the constraints 
\ber \label{constrQuantile}
\tilde \xi_j = \frac{1 - 2\tau_j}{\tau_j(1-\tau_j)}   \qquad \mbox{and}  \qquad \tilde \sigma_j^2=\frac{2}{\tau_j(1- \tau_j)}, \qquad j=1,..., p
\eer
\noindent represent necessary  conditions to guarantee that the model in (\ref{margQm}) holds,  as shown in the proposition below.}

\noindent {
\begin{proposition} \label{MargQuantiles}
Let $\bs Y \sim \mbox{MAL}_p(\bs   \mu,\bs D \bs{\tilde  \xi}, \bs D \bs {\tilde\Sigma} \bs D) $ as defined in (\ref{mixtureALD}), and let $\bs \tau= [\tau_1, \tau_2,..., \tau_p]'$ be a fixed $p \times 1$ vector, such that $\tau_j \in (0,1)$, $j=1,...,p$, 
then, 
\berr
\mathbb{P}(Y_j <\mu_j)= \tau_j
\eerr
\noindent if and only if
\ber \label{constraints}
\tilde \xi_j = \frac{1 - 2\tau_j}{\tau_j(1-\tau_j)}   \qquad \mbox{and}  \qquad \tilde \sigma_j^2=\frac{2}{\tau_j(1- \tau_j)}.
\eer 
\noindent In addition, $Y_j \sim \mbox{AL} (\mu_j, \tau_j, \delta_j)$. 
\end{proposition}
}

\noindent \textit{\textbf{Proof}.  } See Appendix. 

\noindent  {Notice that, the representation in (\ref{MALdensityCons}), under (\ref{constrQuantile}), is a suitable reparameterization of Kotz et al. (2001). Indeed the introduction of  the diagonal matrix $\bs D$ ensures that  each  $\delta_j$ represents the scale parameter of the marginal AL distribution of $Y_j$, for every $j =1,2,..., p$ 
 (see the proof  of  Proposition \ref{MargQuantiles} in Appendix for further details). }

\noindent {It is also worth adding a brief comment on parameters identifiability of the MAL distribution, {as one could reasonably ask whether $\bs{D \tilde{\xi}}$ and $\bs{D \tilde \Sigma D} $ are uniquely identified}. 
In the next proposition, we argue that the constraints (\ref{constrQuantile}) in (\ref{mixtureALD}) are necessary conditions for model identifiability, for any fixed quantile level $\tau_j, j=1,...,p$. 
 }

\noindent {\begin{proposition} \label{identifiability}
Let $\bs Y \in {\cal R}^p$ be distributed as a $\mbox{MAL}_p \left( \bs \mu,\, \bs{D \tilde{\xi}},\,  \bs{D \tilde \Sigma D}  \right)$, where $\bs D= \diag [\delta_1, \delta_2,..., \delta_p]$, $\delta_j>0$,  $ \bs {\tilde\xi}= [\tilde \xi_1, \tilde \xi_2,...,\tilde  \xi_p]'$, with  $\tilde \xi_j= \frac{1- 2 \tau_j}{\tau_j(1 - \tau_j)}$ being known for any fixed value of $\tau_j$. $\bs {\tilde \Sigma}$ is a $p \times p$ positive definite matrix such that $\bs {\tilde \Sigma} = \bs{\tilde \Lambda} \bs \Psi \bs{\tilde \Lambda}$, with $\bs \Psi $ being an unknown correlation matrix and $\bs{\tilde \Lambda}= \diag[\tilde \sigma_1, \tilde \sigma_1,..., \tilde \sigma_p]$, with fixed element $\tilde \sigma_j^2= \frac{2}{\tau_j (1 - \tau_j)}$, for every  $j=1,..., p$. Then, the parameters $\bs D$ and $\bs \Psi$ are uniquely identified. 
\end{proposition}}

\noindent \textit{\textbf{Proof}.  } See Appendix.

\noindent {In other words, the constraints in (\ref{constrQuantile}) represent essential conditions not only to retrieve the joint quantile regression model in (\ref{margQm}) (as discussed in Proposition \ref{MargQuantiles}), but also to ensure that the model does not suffer from identifiability problems (see Proposition \ref{identifiability}). }
\noindent

\section{Maximum Likelihood estimation } \label{EM MLE}

\noindent {As shown in the previous sections, the MAL density represents a convenient  tool to  jointly model marginal conditional quantiles of a multivariate response variable in a quantile regression framework. In this section we introduce a maximum likelihood approach to estimate and  make inference on model parameters. We propose a suitable likelihood-based EM algorithm (Dempster et al, 1977), showing that model parameters can be easily obtained in closed form, hence facilitating the computational burden of the algorithm compared to the direct maximization approach. Moreover, as mentioned in the Introduction, given the possible high dimensionality problem in multivariate settings, we also  propose a penalized version (PEM) of the EM algorithm by considering a LASSO regularization approach (Tibshrani, 1996). The procedure essentially modifies the $M$-step of the EM algorithm by introducing a penalty term and provides a data-driven procedure for variable selection.     }

\subsection{The EM algorithm} \label{EM}
\noindent {In this section we propose a Maximum Likelihood-based approach to estimate the parameters of the quantile regression models defined in (\ref{margQm}).  Specifically, we derive a new EM algorithm by exploiting the Gaussian location-scale mixture representation (\ref{mixtureALD}) of the MAL distribution, under the constraints in (\ref{constrQuantile}).  The EM algorithm essentially  alternates between performing an expectation (E) step, which defines the expectation of the complete log-likelihood function evaluated using the current estimate for the parameters, and a maximization (M) step, which computes parameter estimates by maximizing the expected complete log-likelihood obtained in the E-step. 
The expected complete log-likelihood function and the optimal parameter estimators are given below  in the following two propositions. All the proofs  are collected in the Appendix.}

\noindent {For sake of clarity, in the following  we introduce the  notation   $\bs D_{(\bs \delta)}$ and ${\bs{\tilde \Sigma}}_{(\bs \Psi)}$  to make clear that the matrices $\bs D$ and $\bs{\tilde \Sigma} $ depend on the parameters $\bs \delta= [\delta_1,..., \delta_p]'$ and $\bs \Psi$, respectively. }

\begin{proposition} \label{expectedloglik}
{For a {given vector } $\bs \tau=[\tau_1, \tau_2,...,\tau_p]'$, let $\bs D_{(\bs \delta)}=\diag[\bs \delta]$, with $\bs \delta=[\delta_1,..., \delta_p]'$ and let $\tilde {\bs\Sigma}_{(\bs \Psi)}=\bs{\tilde \Lambda \Psi \tilde \Lambda}$ with $\bs{\tilde \Lambda}$ subject to the  constrains in (\ref{constraints})}. Let $\tilde {m_i}= (\bs y_i- \bs \beta_{\tau}\bs X_i)' (\bs D_{(\bs \delta)} \tilde{ \bs \Sigma}_{(\bs \Psi)} \bs D_{(\bs \delta)})^{-1}(\bs y_i- \bs \beta_{\tau}\bs X_i)$, for every $i=1,2,...,n$, and define  $\tilde d= \tilde {\bs \xi}' \tilde{ \bs \Sigma}_{(\bs \Psi)} \tilde{ \bs \xi}$. Then, the expected complete log-likelihood  function  (up to additive constants) is 

\ber \label{OF}
E\left[l_c(\bs \beta_{\tau}, \bs D_{(\bs \delta)},\tilde {\bs\Sigma}_{(\bs \Psi)}|\bs Y_i, \hat{\bs \beta}_{\tau}, \bs{D}_{(\hat{\bs \delta})}, {\tilde {\bs\Sigma}}_{(\hat{\bs \Psi})})  \right] &=& -\frac{n}{2} \log |\bs D_{(\hat{\bs \delta})} \tilde  {\bs \Sigma}_{(\hat{\bs \Psi})} \bs D_{(\hat{\bs \delta})}| + \sum_{i=1}^n (\bs Y_i - \bs \beta_{\tau}X_i)' \bs D_{(\hat{\bs \delta})}^{-1} \tilde{ {\bs \Sigma}}_{(\hat{\bs \Psi})}^{-1}\tilde {\bs \xi} \nonumber\\[-0.1in]
&&  \label{OF1}\\
& -& \frac{1}{2} \sum_{i=1}^nz_i(\bs Y_i-\bs \beta_{\tau}\bs X_i)' (\bs D_{(\hat{\bs \delta})} \tilde  {\bs \Sigma}_{(\hat{\bs \Psi})} \bs D_{(\hat{\bs \delta})})^{-1} (\bs Y_i - \bs \beta_{\tau}\bs X_i) \nonumber \\ [-0.1in]
&&\label{OF2} \\
&-& \frac{1}{2} \tilde{{\bs \xi}}'\tilde  {\bs \Sigma}_{(\hat{\bs \Psi})} \tilde {\bs \xi}\sum_{i=1}^n u_i  \label{OF3}.
\eer

\noindent where

 \ber \label{wi}
u_i=  \left(  \frac{\hat {\tilde {m}}_i}{2+\hat {\tilde d}}  \right)^{\frac{1}{2}} \frac{K_{\nu +1}\left( \sqrt{(2+\hat{\tilde d})\hat {\tilde {m}}_i} \right)}{K_{\nu}\left(   \sqrt{(2+\hat {\tilde d}) \hat {\tilde {m}}_i}\right)}, \qquad 
 z_i= \left( \frac{2+\hat {\tilde d}}{\hat {\tilde {m}}_i} \right)^{\frac{1}{2}}  \frac{K_{\nu +1}  \left( \sqrt{(2+\hat {\tilde d})\hat {\tilde {m}}_i} \right)}{K_{\nu}  \left( \sqrt{(2+\hat {\tilde d})\hat {\tilde {m}}_i} \right)} - \frac{2 \nu}{\hat {\tilde {m}}_i}  
\eer

%

\noindent with

\ber
\hat {\tilde{m}}_i &=& (\bs y_i- \hat{\bs \beta}_{\tau}\bs X_i)' (\bs D_{(\hat{\bs \delta})} {\tilde {\bs \Sigma}}_{\hat{\bs \Psi}} \bs D_{(\hat{\bs \delta})})^{-1}(\bs y_i- \hat {\bs \beta}_{\tau}\bs X_i) \\
\hat{ \tilde d} &=& \tilde {\bs \xi}' {\tilde {\bs \Sigma} }_{(\hat{\bs \Psi})}\tilde{ \bs \xi}.
\eer
\end{proposition} 

\noindent \textit{\textbf{Proof}.  } See Appendix.

\noindent {For a given $\bs \tau$}, maximizing the expectation of the complete data log-likelihood in (\ref{OF1})-(\ref{OF3}) with respect to the parameters ${\bs \beta}_{\tau}, \bs \Psi$, and $\bs \delta$ leads to the following M-step updates.

\begin{proposition} \label{Mestimates}
{Given the vector} $\bs{\tau}$, the values of ${\bs \beta}_{\tau}, \tilde{\bs{\Sigma}}_{(\bs \Psi)}$ and $ \bs \delta$ maximizing  (\ref{OF1})-(\ref{OF3}) are

\ber 
\hat {\bs \beta}_{\tau} '&=& \left( \sum_{i=1}^n z_i\bs X_i\bs X_i' \right)^{-1} \left( \sum_{i=1}^n z_i\bs X_i\bs Y_i' - \sum_{i=1}^n \bs X_i\tilde{\bs \xi}' \bs D_{(\hat{\bs \delta})} \right) \label{focBeta}\\[0.1in]
{\tilde{ \bs \Sigma}}_{(\hat{\bs{\Psi}})}&=& \frac{1}{n}\sum_{i=1}^n z_i \bs D^{-1}_{(\hat{\bs \delta})}(\bs Y_i - \hat{\bs \beta}_{\tau}\bs X_i) (\bs Y_i - \hat{\bs \beta}_{\tau}\bs X_i)'\bs D^{-1}_{(\hat{\bs \delta})} + \frac{1}{n}\sum_{i=1}^nu_i \tilde{\bs \xi} \tilde {\bs \xi}'
- \frac{2}{n} \bs D^{-1}_{\hat{\bs \delta}}\sum_{i=1}^n(\bs Y_i -\hat{\bs \beta}_{\tau}\bs X_i)' \tilde {\bs \xi}' ,\nonumber \\[-0.01in]
&& \label{focSigma}
\eer 

\noindent while the estimation of  $\boldsymbol{\delta}$ is obtained through a numerical optimization, by solving the following non linear first order condition 

\ber \label{focD}
\sum_{i=1}^n(\bs Y_i - \hat{\bs \beta}_{\tau}\bs X_i)\tilde{\bs \xi}'\tilde{\bs \Sigma}_{(\hat{\bs \Psi})} + n\bs D_{(\hat{\bs \delta})} - \sum_{i=1}^n Z_i (\bs Y_i - \hat{\bs \beta}_{\tau}\bs X_i)(\bs Y_i - \hat{ \bs \beta}_{\tau}\bs X_i)'{\bs D}^{-1}_{(\hat{\bs \delta})}\tilde {\bs \Sigma}_{(\hat{\bs \Psi})}=0_{p \times p}
\eer
\end{proposition}

\noindent \textit{\textbf{Proof}.  } See Appendix.

\noindent Therefore, the EM algorithm can be implemented as follows: 

\medskip

\noindent \textit{E-step}: Set the iteration number $h=1$. {Fix the vector $\bs{\tau}$ at the chosen quantile levels $\tau_1,..., \tau_p$ of interest and } initialize the parameter  $\boldsymbol{\theta}= (\bs{\beta}_{\tau}, \Psi, \boldsymbol{\delta} ) $, deriving $\tilde{\bs {\Sigma}}_{(\bs \Psi)}$ and $\bs{D}_{(\bs \delta)}$. Then, given $\boldsymbol{\theta}= \boldsymbol{\theta}^{(h)}$, calculate  the weights

\ber
u_i^{(h)}&=& \left( \frac{m_i^{(h)}}{2+\tilde d^{(h)}} \right)^{\frac{1}{2}}  \frac{K_{\nu +1} \left( \sqrt{\left(2 + \tilde d^{(h)}\right)\tilde m_i^{(h)}} \right)}{K_{\nu}\left( \sqrt{\left(2 + \tilde d^{(h)}\right) \tilde{m}_i^{(h)}} \right)} \label{weightsu}\\[0.1in]
z_i^{(h)}&=& \left( \frac{2+ {\tilde d^{(h)}}}{{\tilde {m}}_i^{(h)}} \right)^{\frac{1}{2}}  \frac{K_{\nu +1}  \left( \sqrt{\left((2+ {\tilde d}^{(h)}\right) {\tilde {m}}_i^{(h)}} \right)}{K_{\nu}  \left( \sqrt{\left(2+{\tilde d^{(h)}}\right) {\tilde {m}}_i^{(h)}} \right)} - \frac{2 \nu}{ {\tilde {m}}_i^{(h)}}  \label{weightsz}
\eer

\noindent where $\tilde{m_i}^{(h)}=\left(\bs Y_i - \bs \beta_{\tau}^{(h)}\bs X_i  \right) \left( \bs D^{(h)}_{(\bs \delta)}\tilde{\bs \Sigma}^{(h)}_{(\bs \Psi)}\bs D^{(h)}_{(\bs \delta)} \right)^{-1}\left(\bs Y_i -\bs \beta_{\tau}^{(h)}\bs X_i\right) $, and $\tilde{d}^{(h)}=\tilde{\bs \xi}' \tilde{\bs \Sigma}^{(h)}_{(\bs \Psi)}\tilde{\bs  \xi}$, for $i =1,2,.., n$.

\noindent \textit{M-step}: Use $u_i^{(h)}$ and $z_{i}^{(h)}$ to maximize $E[l_c(\boldsymbol{\theta} | \boldsymbol{\theta}^{(h)}]$ with respect to $\boldsymbol{\theta}$, and obtain the new parameter estimates as

\ber \label{betaMstep}
\hat{\boldsymbol{\beta}}_{\tau}^{'(h+1)}&=& \left( \sum_{i=1}^n z_i^{(h)}\bs X_i\bs X_i' \right)^{-1} \left( \sum_{i=1}^n z_i^{(h)}\bs X_i\bs Y_i' - \sum_{i=1}^n X_i\tilde{\bs\xi}' \bs D^{(h)}_{(\hat{\bs{\delta}})} \right) \label{Beta}\\[0.1in]
\boldsymbol{\tilde \Sigma}^{(h+1)}_{(\hat{\bs \Psi})}&=&\frac{1}{n}\sum_{i=1}^n z_i^{(h)}\bs D^{-1(h)}_{(\hat{\bs \delta})}(\bs Y_i - \hat{\bs \beta}_{\tau}^{(h+1)}\bs X_i) (\bs Y_i - \hat{\bs \beta}_{\tau}^{(h+1)}\bs X_i)'\bs D^{-1(h)}_{(\hat{\bs \delta})}\label{Sigma1}\\
& +& \frac{1}{n}\sum_{i=1}^nu_i^{(h)} \tilde{\bs \xi} \tilde{\bs \xi}'
- \frac{2}{n}  \bs D^{-1(h)}_{(\hat{\bs \delta})}\sum_{i=1}^n(\bs Y_i -\hat{\bs \beta}_{\tau}^{(h+1)}\bs X_i)' \tilde {\bs \xi}' \label{Sigma2} \label{SigmaMstep}.
\eer

 \noindent while $\boldsymbol{D}^{(h+1)}_{(\bs \delta)}$ is obtained as the solution of the following equation

\ber \label{D}
\sum_{i=1}^n(\bs Y_i - \hat{\bs \beta}_{\tau}^{(h+1)}\bs X_i)\tilde{\bs\xi}'\tilde{\bs \Sigma}^{(h+1)}_{(\hat{\bs \Psi})} + n\bs D_{(\bs \delta)} - \sum_{i=1}^n z_i^{(h)} (\bs Y_i - \hat{ \bs \beta}_{\tau}^{(h+1)}\bs X_i)(\bs Y_i - \hat{\bs \beta}_{\tau}^{(h+1)}\bs X_i)'\bs D^{-1}_{(\bs \delta)}\tilde{ \bs \Sigma}^{(h+1)}_{(\hat{\bs \Psi})}=0_{p \times p} \label{DMstep}
\eer

\noindent The procedure is iterated until convergence, {that is when the difference between the likelihood function evaluated at two consecutive iterations is small enough\footnote{In our paper we set this convergence criterion equal to $10^{-6}$.}. }

\noindent {Notice that all the parameter estimates in (\ref{betaMstep})-(\ref{DMstep}) account for the multivariate structure of the data through the weights $u_i$ and $z_i$ in (\ref{weightsu}) and (\ref{weightsz}), which depend on the index $\nu$ (as a function of $p$). In the univariate case, $\nu= \frac{1}{2}$ and  the estimators reduce to the case of Sanchez et al. (2013).  }


\subsection{Variable selection and the Penalized EM (PEM) algorithm} \label{PEM}
\noindent {As mentioned before, when dealing with high dimensional statistical problems,  Lasso penalized procedures represent possible solutions to detect significant predictors from a large pool of candidate variables. Therefore, in this section we introduce a penalized version of the EM  algorithm described in Section \ref{EM}}.  The PEM algorithm was originally proposed by Green (1990) to allow for the maximization of a difficult-to-calculate penalized likelihood. Compared to the EM, the PEM algorithm leaves the $E$-step unchanged and modifies the $M$-step with a penalty function introduced to achieve shrinkage (e.g. ridge regression, Hoerl and Kennard, 1970), variable selection (e.g. Lasso, Tibshirani (1996)) or simultaneous shrinkage and variable selection (e.g elastic net, Zou and Hastie (2005)). 
{For a chosen level $\bs \tau$, let us denote by $\bs \theta =\left(\bs \beta_{\tau}, \bs D_{(\bs \delta)},\tilde {\bs \Sigma}_{\bs \Psi} \right)$ the parameter set. Then,  consider the following penalized-$M$ (PM) step:}

\ber \label{thetaPEM}
\bs \theta^{(h+1)}&=& \underset{\bs \theta}{\argmax} \left( { Q}(\bs \theta | \bs \theta^{(h)}) - \lambda \sum_{j=1}^p \sum_{s=1}^k |\beta_{js, \tau_j}|\right)
\eer

\noindent where
\ber \label{OF}
{ Q}(\bs \theta | \bs \theta^{(h)}) &=& -\frac{n}{2} \log |\bs D_{(\bs \delta)} \tilde  {\bs \Sigma}_{(\bs \Psi)} \bs D_{(\bs \delta)}| + \sum_{i=1}^n (\bs Y_i - \bs \beta_{\tau}X_i)' \bs D_{(\bs \delta)}^{-1} \tilde{ {\bs \Sigma}}_{(\bs \Psi)}^{-1}\tilde {\bs \xi} \nonumber\\[-0.1in]
&&  \label{OF1a}\\
& -& \frac{1}{2} \sum_{i=1}^nz_i(\bs Y_i-\bs \beta_{\tau}\bs X_i)' (\bs D_{(\bs \delta)} \tilde  {\bs \Sigma}_{(\bs \Psi)} \bs D_{(\bs \delta)})^{-1} (\bs Y_i - \bs \beta_{\tau}\bs X_i) \label{OF2a}\\
&-& \frac{1}{2} \tilde{{\bs \xi}}'\tilde  {\bs \Sigma}_{(\bs \Psi)} \tilde {\bs \xi}\sum_{i=1}^n u_i  \label{OF3a}.
\eer

\noindent {with all the quantities defined as in Section \ref{EM}. In (\ref{thetaPEM}), the quantity}{ $\lambda \sum_{j=1}^p \sum_{s=1}^k |\beta_{js,\tau_j}|$ represents a convex penalty function, where  $\lambda$ is a tuning parameter that regulates the strength of the penalization assigned to the coefficients in the model.  The choice of $\lambda$ is made by performing cross-validation
 techniques which allows us to consider $\lambda$ as a data-driven parameter. }
In particular, we compute the solutions for a decreasing sequence of values for $\lambda$, starting from the smallest value  $\lambda_{max}$ for which the entire vector $\hat{\bs \beta}_k=0$. We then select a minimum value  $\lambda_{min}=\epsilon \lambda_{max} $ and construct a sequence of $m$ values of $\lambda$, decreasing from $\lambda_{max}$ to $\lambda_{min}$ on the log scale.  This would lead to a more stable algorithm, see e.g. Friedman et al. (2010).  

\noindent Notice that, even though the algorithm penalizes only the M-step, the expressions of the weights $u_i$ and $z_i$ in the E-step will be indirectly affected by the new (penalized) estimates in the M-step at each iteration. 

\noindent In the next section we assess the performance of the EM and PEM algorithms using a simulation exercise.
\section{Simulation study} \label{Simulation}


\noindent {In the following sections we conduct a simulation study to evaluate the small sample properties of the proposed methods. The idea of this exercise is to show that both the EM and the PEM algorithms represent valid procedures to estimate the quantile regression coefficients, regardless of the true data generation process.  
}


\subsection{Joint quantile regression}
\noindent{In this section we asses  the performance of our estimation procedure by simulating a joint quantile regression model as described in Section \ref{MultQuantRegressionMAL}. 
For this purpose, we consider a simple case of $n=1000$ units, of dimension $p=3$ and two explanatory variables.  The observations are generated using the following data generating process:}

\ber \label{modelSimul}
\bs Y_i = \bs \beta_{\tau,0}+ \bs \beta_{\tau,1} X_{i2} + \bs \beta_{\tau,2} X_{i3} + \bs \epsilon_i\qquad i=1,2,...,n
\eer

\noindent The  covariates are randomly drawn from a standard Normal distribution. 
The true value of $\bs \beta_{\tau}$ 
is set equal to \ber\bs \beta_{\tau}= \bqmatrix   -0.382 & -0.372 & 0.715\\
                                     1.993 &0.650 & 0.764\\
                                     0.670 & 1.079 & 0.584 \eqmatrix.\label{betatrue}\eer
We analyze three different quantile vectors. In the first case, we assume  $\bs{\tau}=[0.50, 0.50.0.50]'$, which implies that $\tilde{\bs{\xi}}= [0,0,0]'$ and $\tilde{\bs \Lambda}= \mbox{diag}(2.828,2.828,2.828) $. In the second scenario, we set  $\bs{\tau}=[0.25, 0.50, 0.75]'$ and, consequently, $\tilde{\bs{\xi}}= [2.667,0,-2.667]'$ and $\tilde{\bs \Lambda}= \mbox{diag}(3.266,2.828,3.266) $. Finally, we consider a more extreme case with $\bs{\tau}=[0.90, 0.50, 0.10]'$, $\tilde{\bs{\xi}}= [8.889,0,-8.889]'$ and $\tilde{\bs \Lambda}= \mbox{diag}(4.714,2.828,4.714)$ . For each of the three cases,  the parameter vector is represented by $\bs{\theta}= [\bs{\beta}_{0},\,  \bs{\beta}_{1},\,  \bs{\beta}_{2},\,  \delta_1,\,  \delta_2,\,  \delta_3, \, \rho_{12}, \, \rho_{13}, \, \rho_{23}]$,  where the true values of $\bs{\beta}_{0},\,  \bs{\beta}_{1},\,  \bs{\beta}_{2},$ are defined by the columns of $\bs \beta_{\tau}$ in (\ref{betatrue}) and where we set $\delta_1=0.13, \, \delta_2=0.30\, , \delta_3=0.23$, with $\rho_{12}=0.50\, , \rho_{13}=0.30$ and $\rho_{23}=0.40$.  

\noindent {Two different distributions for the error term generating process are considered in each simulation study:
(a) a multivariate Normal random variable with zero mean and a variance-covariance matrix equal to  $(\bs{D \tilde \xi \tilde \xi' D} +  \bs{D \tilde \Sigma D} )$,  and (b) a multivariate Student $t$ distribution with 3 degrees of freedom, scale parameter $\bs{D \tilde \Sigma D}$ and non centrality parameter equal to $\bs{D \tilde \xi}$.
\noindent  For each distribution of the error term, we carry out $B=500$ Monte Carlo replications and report the relative bias and the root mean square error (RMSE), averaged across the 500 simulations, for each parameter value in $\bs \theta$. The results are shown in Tables \ref{t1} and  \ref{t1b}.} 

\noindent {Table \ref{t1} analyzes the regression coefficient estimates for  the three quantile levels described in the Panel A, Panel B and Panel C. For each of the three panels, the two error term distributions are considered and the corresponding point estimates and percentage bias are reported. As can be easily inferred,  the bias effect is quite small when we analyze the median levels  (see Panel A). As the quantile levels become more extreme (see Panels B and C), the bias slightly increases but it still  remains reasonably small.} 

\noindent {In Table \ref{t1b} we report the Root Mean Sqaure Error (RMSE) of the regression coefficients for the same $\bs \tau$-levels under the same error distributions of Table \ref{t1}, and compare the results obtained by running both the proposed joint quantile regression (\textit{Joint QR RMSE} in the table) and the univariate quantile regressions (\textit{Univariate QR RMSE} in the table) separately for each  marginal $Y_j$. In this way we want to highlight the added value of using the MAL distribution for multiple quantile regression purpose, which accounts for potential correlation among the responses.}
\noindent {It is worth noting  that in all the simulation situations, the effective gain of the proposed joint approach  is reflected in the smaller RMSE of the estimates compared to those in the univariate case.}

\subsection{Penalized joint quantile regression}

\noindent {In this section, a simulation study is proposed to evaluate the performance of the  penalized joint quantile regression, which uses  the PEM algorithm proposed in Section \ref{PEM}.}
We analyze a simple case with $n=1000$, $p=3$ and a set of four explanatory variables using the same data generating process of the previous section. {For a fixed $\bs \tau$-level, the matrix of regression coefficients $\bs \beta_{\tau}$ 
contains 15 elements, where we set six of them (namely, $\beta_{12}, \beta_{14}, \beta_{23}, \beta_{24}, \beta_{32}, \beta_{33}$) equal to zero.  
As in the previous section, we analyze three different quantile vectors, i.e. $\bs{\tau}=[0.50, 0.50, 0.50]'$, $\bs{\tau}=[0.25,0.50,0.75]'$, and $\bs \tau=[0.10,0.50, 0.90]$. For each of the three cases we perform 100 Monte Carlo simulations, under either  the ${\cal N}_3(\bs 0, \bs D \bs{\tilde \xi} \bs{\tilde \xi}' \bs D +  \bs{D \tilde \Sigma D})$ and the $t_3(\bs{D \tilde \Sigma D}, \bs{D \tilde \xi})$ distributions as possible data generating process.   
Then, for each case,  we estimate the model parameters using the penalized objective function defined in  (\ref{thetaPEM}). The estimation of the tuning parameter $\lambda$ is obtained using a 10-fold cross validation method, where the initial grid of the possible values for $\lambda \in [\lambda_{min}, \lambda_{max}]$ has been  described in Section \ref{PEM}.} 

\noindent Table \ref{t2lasso} reports the true positive rate (TPR) for each of the true coefficients initially set equal to zero. The TPR gives a measure of how sensitive a given method is at discovering non-zero entries and we calculate it as the ratio between the number of simulations that correctly identify the parameter as a zero value, over the total number of simulations (i.e. the number of true zeros for each coefficients).   

\noindent The results show that the PEM method performs quite well, with an average TPR of more than the 80\% across the three simulation scenarios and regardless of the distributional assumption of the errors. 

\section{Empirical application} \label{Empirical application}

\noindent {As stated in the Introduction,  recent financial and economic crises have put the need for a thorough analysis of the causes and the effects of the financial distress. {In this context, quantile regression has turned out to be an effective framework to study and evaluate financial stability of systems.} 
{In line with the recent literature that links quantile regression with  measures of financial risks ({see for example Adrian
and  Brunnermeier (2016),  Covas et al. (2014), Engle and Manganelli (2004) and  Xiao et al. (2015)}), in this section we use our methodology to identify the main determinants for a risk of financial distress.}  We use data on 2,020 private limited non-financial Italian firms from the Amadeus Bureau van Dijk dataset, with reference year 2015. Following Pindado et al. (2008) and Bastos and Pindado (2013) among others, we adopt a definition of financial distress that evaluates the firm's capacity to satisfy its financial obligations. {Specifically, 
they classify  a company as
financially distressed not only when it files for bankruptcy but also when the two following events occur: (1) its
earnings before interest and taxes depreciation and amortization
(EBITDA) is under the first quartile of the sample or  (2) the firm's leverage is above the
third quartile of the sample.} 

\noindent {The idea of this real data application is to use our joint estimation approach to study the entire distribution of both firm's leverage and EBITDA  and assess how the impact of firm's characteristics (such as  profitability, financial expenses and earnings) varies with different quantile levels. 
This allows us to identify not only the main determinants of firm's risk of financial distress, but also how the effect of these factors may vary depending on the severity of the distress a firm is facing.}

 
\noindent As a measure of firm's leverage (\textit{leverage}) we use the ratio between firm's total asset and equity. As explanatory variables, we consider indexes of firm's profitability (\textit{profit}), financial expenses (\textit{finexp}), and retained earnings (\textit{earnings}), as suggested in  Pindado et al. (2008).  We also consider the impact of short-term debt (\textit{current debt}) and the ratio between firm's cashflow and total assets ($\textit{cashflow}$), and control  for other firm's specific characteristics such as firms' total fixed assets over total asset ratio (\textit{fixassets}, which can be interpreted as an indicator of firm's collaterals),  firm's net income scaled by total assets (\textit{netincome}, as a proxi of the activity level of the firm), and size, measured by the number of employees in the firm (\textit{employees}).   

\noindent The results for the two cases when $\bs{\tau}=[0.75, 0.25]'$  and $\bs{\tau}=[0.90, 0.10]'$  are reported in Table \ref{t3}, Panel A and Panel B, respectively. {In the table, parameter estimates are displayed in boldface when significant at the standard 5\% level. Standard errors of the estimates are computed using non parametric bootstrap (see e.g., Geraci and Bottai (2007)) and are reported in brackets. }

\noindent We find a negative relationship between profitability and financial distress, as the coefficient of profitability (\textit{profit}) shows  a significant and negative impact on both EBITDA and leverage. This is in line with the results in    Pindado et al. (2008) and Campbell et al. (2008), who argue that  firms that
face financial distress are most likely unable to fulfill their
financial obligations. This effect remains constant also at more extreme quantile levels of the distribution, as shown in Panel B of Table \ref{t3}.

\noindent We also find a positive effect of financial expenses, whose magnitude is amplified especially for the leverage component. This effect is still consistent with the findings in Pindado et al. (2008) and confirms the expectations that the risk of financial distress increases as the firm's risk of not
being able to comply with its financial obligations rises. This effect captures the firm's
financial vulnerability, which  increases when considering more extreme (riskier) values of leverage. 

\noindent In analyzing the effect of retained earnings on financial distress,  we find evidence of a positive impact on both EBITDA and leverage, even though it decreases as  riskier quantiles are analyzed. This is in contrast with the expected results in the literature, where a negative relationship with financial distress likelihood is  typically postulated, as a firm should have a lower capacity of self-financing during periods of higher financial stress. However, several papers have documented the relationship between financial distress and earnings management practices in economy like Italy, and find evidence that private companies experiencing financial distress tend to manipulate their earnings to portray better financial performance and obtain bank financing (see e.g Bisogno and De Luca (2015), among others). In this case, earnings should not be considered a very informative determinant of firm's financial distress.

\noindent Another important factor in explaining financial distress is the ratio between firm's fixed asset over total asset. The claim is typically that  tangible assets  tend
to reduce the financial distress costs because of the liquidation possibility in case of default (see e.g Charalambakis and Psychoyios (2012)). Hence, the higher is the risk of a financial distress, the higher is the level of tangible assets a firm will have on its balance sheet.  Our results confirm this evidence, showing a positive  effect of \textit{fixasset} on both EBITDA and leverage. 

\noindent Firm's cash flow over total asset ratio is also a significant predictor of financial distress. It's impact on leverage is highly negative, whilst it is found to be positive on the EBITDA component, with a sensible decrease when moving towards more extreme quantile.  This is could be due to the fact that firms with  low cash flow are less likely to make leverage adjustments, especially during period of financial stress. Predictive power is also shown by the firm's short-term debt. Finally, no significant effect is found in terms of size (measured by the number of employees) and activity level (\textit{netincome}). 

\noindent {In Table \ref{t4} we report the LASSO estimates {computed on the same firms'  sample and on the same quantile levels}, which essentially confirm the above findings.}


\section{Conclusions} \label{Conclusions}

\noindent {This paper proposes a new likelihood-based method to jointly estimate marginal conditional quantiles of a multivariate response variable in a linear regression framework. We use a suitable reparameterization of the Multivariate Asymmetric Laplace distribution of Kotz et al. (2001), whose  mixture representation allows us to implement a new EM algorithm. Using this procedure, we show that the regression parameters can be easily estimated in closed form, hence avoiding direct maximization procedures.  A penalized version of the algorithm is also proposed as an automatic data-driven procedure to perform variable selection. } 
The good performance of the two methods is evaluated using a simulation exercise, where  extreme quantiles  are also considered as possible simulation scenarios. 
An empirical application to financial distress analysis on a sample of Italian firms is finally presented.} 

\noindent {As the approach of quantile regression is widely used in the literature, several  extensions of the results obtained in this paper can be analyzed in future research. Longitudinal settings and/or random effects models are of particular interest in this context, with the goal of characterizing the change in the response variables  over time and accounting also for the dependence between serial observations on the same subject. Other forms of penalized algorithm could be also promising, as the simultaneous regularization of the parameters could perform better when a large set of (possibly correlated) explanatory variables  is used in the application. }

\clearpage

\begin{table}[h!]
\linespread{1}
\par
\footnotesize
\caption{\textbf{Parameter estimates at different quantile levels: point estimates and relative bias.} 
 }
\label{t1}
\begin{center}
\begin{tabular}{lccccccccccccccccccccccccccccccccc}
\hline \hline\\[0.005in]
& \multicolumn{1}{c}{$\hat{\beta}_{01}$}& \multicolumn{1}{c}{$\hat{\beta}_{02}$}& \multicolumn{1}{c}{$\hat{\beta}_{03}$}& &
\multicolumn{1}{c}{$\hat{\beta}_{11}$}& \multicolumn{1}{c}{$\hat{\beta}_{12}$}& \multicolumn{1}{c}{$\hat{\beta}_{13}$} && 
\multicolumn{1}{c}{$\hat{\beta}_{21}$}& \multicolumn{1}{c}{$\hat{\beta}_{22}$}& \multicolumn{1}{c}{$\hat{\beta}_{23}$}\\[0.05in]
\hline\\
& \multicolumn{11}{c}{Panel A: $\bs \tau=[0.50,0.50,0.50]'$}\\[0.05in]
\multicolumn{1}{l}{$\bs{\epsilon}_i \sim {\cal N}_3(\bs 0, \bs{D \tilde \xi \tilde \xi' D} +  \bs{D \tilde \Sigma D} )$}&&&&&&&\\[0.05in]
Estimate &   -0.382&1.984&0.667&&-0.374&0.653&1.083&&0.713&0.771&0.586\\
Bias (\%)&    -0.063&-0.423&-0.325&&0.560&0.443&0.423&&-0.199&0.871&0.305\\[0.1in]
\multicolumn{1}{l}{$\bs{\epsilon}_i \sim t_3(\bs{D \tilde \Sigma D}, \bs{D \tilde \xi})$}&&&&&&&\\[0.05in]
Estimate &    -0.379&1.980&0.666&&-0.369&0.646&1.072&&0.706&0.756&0.508\\
Bias (\%)&    0.632&0.105&0.698&&0.749&0.367&0.452&&0.778&0.056&0.898\\[0.1in]

& \multicolumn{11}{c}{Panel B: $\bs \tau=[0.25,0.50,0.75]'$}\\[0.05in]
\multicolumn{1}{l}{$\bs{\epsilon}_i \sim {\cal N}_3(\bs 0, \bs{D \tilde \xi \tilde \xi' D} +  \bs{D \tilde \Sigma D} )$}&&&&&&&\\[0.05in]
Estimate &    -0.394&1.992&0.697&&-0.371&0.647&1.083&&0.714&0.768&0.590\\
Bias (\%)&  3.141&-0.030&4.030&&-0.166&-0.478&0.405&&-0.175&0.609&1.189\\[0.1in]
\multicolumn{1}{l}{$\bs{\epsilon}_i \sim t_3(\bs{D \tilde \Sigma D}, \bs{D \tilde \xi})$}&&&&&&&\\[0.05in]
Estimate &   -.0418&2.023&0.709&&-0.373&0.655&1.086&&0.714&0.764&0.589\\
Bias (\%)&    7.515&1.501&4.776&&0.337&0.817&0.664&&0.510&0.448&0.956\\[0.1in]

& \multicolumn{11}{c}{Panel C: $\bs \tau=[0.10,0.50,0.90]'$}\\[0.05in]
\multicolumn{1}{l}{$\bs{\epsilon}_i \sim {\cal N}_3(\bs 0, \bs{D \tilde \xi \tilde \xi' D} +  \bs{D \tilde \Sigma D} )$}&&&&&&&\\[0.05in]
Estimate &    -0.431&2.006&0.731&&-0.363&0.651&1.065&&0.771&0.764&0.576\\
Bias (\%)&   12.821&0.669&9.104&&-1.413&0.181&-1.218&&-0.539&0.051&-1.271\\[0.1in]
\multicolumn{1}{l}{$\bs{\epsilon}_i \sim t_3(\bs{D \tilde \Sigma D}, \bs{D \tilde \xi})$}&&&&&&&\\[0.05in]
Estimate &    -0.436&2.012&0.755&&-0.386&0.659&1.100&&0.717&0.764&0.590\\
Bias (\%)&    14.222&0.974&12.694&&2.748&1.407&1.987&&1.013&0.435&0.987\\[0.1in]
\hline
\hline
\end{tabular}
\end{center}
\end{table}

\begin{table}[h!]
\linespread{1}
\par
\footnotesize
\caption{\textbf{Joint  quantile estimation  versus univariate quantile regressions: RMSEs.} 
}
\label{t1b}
\begin{center}
\begin{tabular}{lccccccccccccccccccccccccccccccccc}
\hline \hline\\[0.005in]
& \multicolumn{1}{c}{$\hat{\beta}_{01}$}& \multicolumn{1}{c}{$\hat{\beta}_{02}$}& \multicolumn{1}{c}{$\hat{\beta}_{03}$}& &
\multicolumn{1}{c}{$\hat{\beta}_{11}$}& \multicolumn{1}{c}{$\hat{\beta}_{12}$}& \multicolumn{1}{c}{$\hat{\beta}_{13}$} && 
\multicolumn{1}{c}{$\hat{\beta}_{21}$}& \multicolumn{1}{c}{$\hat{\beta}_{22}$}& \multicolumn{1}{c}{$\hat{\beta}_{23}$}\\[0.05in]
\hline\\
& \multicolumn{11}{c}{Panel A: $\bs \tau=[0.50,0.50,0.50]'$}\\[0.05in]
\multicolumn{1}{l}{$\bs{\epsilon}_i \sim {\cal N}_3(\bs 0, \bs{D \tilde \xi \tilde \xi' D} + \bs{D \tilde \Sigma D} )$}&&&&&&&\\[0.05in]
Joint QR RMSE&         0.022 & 0.046 & 0.038 && 0.024 & 0.049& 0.040 && 0.020 & 0.048 & 0.032\\[0.01in]
Univariate QR RMSE&         0.023 & 0.049 & 0.039 && 0.029 & 0.053& 0.041 && 0.020 & 0.049 & 0.034\\[0.1in]
\multicolumn{1}{l}{$\bs{\epsilon}_i \sim t_3(\bs{D \tilde \Sigma D}, \bs{D \tilde \xi})$}&&&&&&&\\[0.05in]
Joint QR RMSE&         0.028 & 0.060 & 0.050 && 0.027 & 0.067& 0.049 && 0.030 & 0.051 & 0.034\\[0.01in]
Univariate QR RMSE&         0.034 & 0.074 & 0.058 && 0.030 & 0.077& 0.044 && 0.027 & 0.067 & 0.059\\[0.2in]

& \multicolumn{11}{c}{Panel B: $\bs \tau=[0.25,0.50,0.75]'$}\\[0.05in]
\multicolumn{1}{l}{$\bs{\epsilon}_i \sim {\cal N}_3(\bs 0, \bs{D \tilde \xi \tilde \xi' D} +  \bs{D \tilde \Sigma D} )$}&&&&&&&\\[0.05in]
Joint QR RMSE&         0.467& 0.049 & 0.283 && 0.029 & 0.049& 0.049 && 0.031 & 0.044 & 0.049\\[0.01in]
Univariate QR RMSE&         0.470 & 0.055 & 0.324 && 0.029 & 0.049& 0.051 && 0.038 & 0.048 & 0.051\\[0.1in]
\multicolumn{1}{l}{$\bs{\epsilon}_i \sim t_3(\bs{D \tilde \Sigma D}, \bs{D \tilde \xi})$}&&&&&&&\\[0.05in]
Joint QR RMSE&         0.428 & 0.059 & 0.292 && 0.053 & 0.049& 0.096 && 0.056 & 0.106 & 0.041\\[0.01in]
Univariate QR RMSE&         0.491 & 0.061 & 0.572 && 0.056 & 0.049& 0.103 && 0.061 & 0.196 & 0.049\\[0.2in]

& \multicolumn{11}{c}{Panel C: $\bs \tau=[0.10,0.50,0.90]'$}\\[0.05in]
\multicolumn{1}{l}{$\bs{\epsilon}_i \sim {\cal N}_3(\bs 0, \bs{D \tilde \xi \tilde \xi' D} +  \bs{D \tilde \Sigma D} )$}&&&&&&&\\[0.05in]
Joint QR RMSE&        0.981 & 0.055 & 0.835 && 0.063& 0.039& 0.130 && 0.062 & 0.045 & 0.131\\[0.01in]
Univariate QR RMSE&         1.063 & 0.05 & 3.274 && 0.087 & 0.048& 0.140 && 0.075 & 0.047 & 0.142\\[0.1in]
\multicolumn{1}{l}{$\bs{\epsilon}_i \sim t_3(\bs{D \tilde \Sigma D}, \bs{D \tilde \xi})$}&&&&&&&\\[0.05in]
Joint QR RMSE&        0.997 & 0.101 & 0.936 && 0.093 & 0.241& 0.165 && 0.065 & 0.048 & 0.152\\[0.01in]
Univariate QR RMSE&         0.692& 0.135 & 1.228 && 0.097 & 0.243& 0.182 && 0.071 & 0.053 & 0.169\\[0.1in]
\hline
\hline
\end{tabular}
\end{center}
\end{table}

\begin{table}[h!]
\linespread{1.3}
\par
\footnotesize
\caption{\textbf{The performance of the PEM algorithm: True Positive Rate (TPR).} 
}
\label{t2lasso}
\setlength{\tabcolsep}{14pt}
\begin{center}
\begin{tabular}{lcccccc}
\hline \hline\\[0.005in]
& \multicolumn{1}{c}{$\hat{\beta}_{12}$}& \multicolumn{1}{c}{$\hat{\beta}_{14}$}& \multicolumn{1}{c}{$\hat{\beta}_{23}$}& 
\multicolumn{1}{c}{$\hat{\beta}_{24}$}& \multicolumn{1}{c}{$\hat{\beta}_{32}$}& \multicolumn{1}{c}{$\hat{\beta}_{33}$} \\[0.05in]
\hline\\
\multicolumn{1}{l}{$\bs{\epsilon}_i \sim{\cal N}_3(\bs 0, \bs{D \tilde \xi \tilde \xi' D} +  \bs{D \tilde \Sigma D} )$}&&&&\\[0.05in]
 $\bs \tau=[0.50,0.50,0.50]'$ &   89.122&92.475&84.341&89.342&91.753&89.134\\[0.05in]
 $\bs \tau=[0.25,0.50,0.75]'$&    87.221&83.453&79.978&86.397&83.641&87.550\\[0.05in]
 $\bs \tau=[0.10,0.50,0.90]'$&         83.341 & 82.512 & 80.123 & 86.101 & 82.308& 86.761\\[0.2in]
\multicolumn{1}{l}{$\bs{\epsilon}_i \sim  t_3(\bs{D \tilde \Sigma D}, \bs{D \tilde \xi})$}&&&&\\[0.05in]
 $\bs \tau=[0.50,0.50,0.50]'$ &   88.324&89.546&84.209&86.008&88.331&88.198\\[0.05in]
 $\bs \tau=[0.25,0.50,0.75]'$&    86.130&83.346&80.001&87.121&82.453&85.345\\[0.05in]
 $\bs \tau=[0.10,0.50,0.90]'$&         83.007 & 81.978 & 81.321 & 86.004 & 80.423& 84.121\\[0.1in]
\hline
\hline
\end{tabular}
\end{center}
\end{table}

\bigskip

\bigskip

\begin{table}[h!]
\linespread{1.5}
\par
\footnotesize
\caption{\textbf{Estimated coefficients at different quantile levels and standard errors.}
}
\label{t3}
\setlength{\tabcolsep}{12pt}
\begin{center}
\begin{tabular}{lrrcrr}

\hline \hline\\
& \multicolumn{1}{c}{Leverage }& \multicolumn{1}{c}{EBITDA } && \multicolumn{1}{c}{Leverage }& \multicolumn{1}{c}{EBITDA} \\[0.01in]
\hline\\
 & \multicolumn{2}{c}{Panel A: $\bs \tau=[0.75,0.25]'$} & & \multicolumn{2}{c}{Panel B: $\bs \tau=[0.90,0.10]'$} \\[0.1in]
Constant &    \textbf{-4.472} (0.501)&\textbf{-0.321} (0.105)&&\textbf{-4.359} (0.557)  &   \textbf{-0.274}  (0.087)\\[0.1in]
profit &    \textbf{-0.022} (0.003)&\textbf{-0.024}  (0.001)&&\textbf{-0.022} (0.003) &    \textbf{-0.024}   (0.005)\\[0.1in]
finexp &    \textbf{13.598}  (0.538)&\textbf{0.451} (0.107)&&\textbf{15.849}  (0.567)&    \textbf{0.312}  (0.091)   \\[0.1in]
earnings &    \textbf{1.118}  (0.509)&\textbf{0.748} (0.110)& &\textbf{0.984} (0.571) &    \textbf{0.621}  (0.092)  \\[0.1in]
employee &    {-0.006}   (0.005)&{-0.001} (0.001)&&{-0.003}  (0.005)&    \textbf{-0.002}  (0.001) \\[0.1in]
fixasset &    \textbf{2.445} (0.132)&\textbf{0.187} (0.027)&&\textbf{2.907} (0.119)&    \textbf{0.151} (0.020)  \\[0.1in]
netincome &   {-0.371}  (1.654)&{0.347} (0.337)& &\textbf{-3.816} (1.658) &    {0.345}   (0.245) \\[0.1in]
current debt &    \textbf{0.107}  (0.008)&\textbf{-0.009} (0.002)&&\textbf{0.122} (0.008)&    \textbf{-0.007}  (0.001)  \\[0.1in]
cashflow &    \textbf{-4.089} (1.376)&\textbf{6.251} (0.251)&&\textbf{-4.704} (1.256)&    \textbf{4.407}   (0.212)\\[0.1in]
\hline\\
$\delta_j$ & \textbf{0.952}  (0.018)&\textbf{0.190} (0.003)  && \textbf{0.800}  (0.012)& \textbf{0.138} (0.002)\\[0.01in]
$\rho_{12}$ &&\textbf{-0.131} (0.027)  && -0.005 (0.032)\\[0.01in]
n&2,020&2,020&&2,020&2,020\\
\hline
\hline
\end{tabular}
\end{center}
\end{table}

\begin{table}[h!]
\linespread{1.3}
\par
\footnotesize
\caption{\textbf{LASSO parameter estimates.} 
}
\label{t4}
\setlength{\tabcolsep}{16pt}
\begin{center}
\begin{tabular}{lrrcrr}

\hline \hline\\
& \multicolumn{1}{c}{EBITDA }& \multicolumn{1}{c}{Leverage } && \multicolumn{1}{c}{EBITDA }& \multicolumn{1}{c}{Leverage} \\[0.01in]
\hline\\
 & \multicolumn{2}{c}{Panel A: $\bs \tau=[0.75,0.25]'$} & & \multicolumn{2}{c}{Panel B: $\bs \tau=[0.90,0.10]'$} \\[0.1in]
Constant &   -3.085 &-0.408 &&-4.566  & -0.364\\[0.1in]
profit &   -0.012&-0.343&&-0.014&  -0.038\\[0.1in]
finexp &    12.624&0.019&&16.222&   0.014  \\[0.1in]
earnings &  0.478&1.422& &1.376&   0.389  \\[0.1in]
employee &   --&--&&--&   --  \\[0.1in]
fixasset &   2.038&0.307&&1.067&  0.235  \\[0.1in]
netincome &   -0.676&--& &-1.267 &   -- \\[0.1in]
current debt &    0.721&-0.330&&0.220&-0.107\\[0.1in]
cashflow &   -3.035&5.606&&-2.072&   5.236\\[0.1in]
\hline\\
$\delta_j$ &0.937  &0.831  &&0.827&0.135\\[0.01in]
$\rho_{12}$ &-0.114&&&-0.006&\\[0.01in]
$\lambda$&2.259&&&5.612&\\[0.01in]
n&2,020&2,020&&2,020&2,020\\
\hline
\hline
\end{tabular}
\end{center}
\end{table}

\clearpage
\section*{Appendix. Proofs of Propositions }

\medskip 

\noindent \textbf{Proof of Proposition \ref{MargQuantiles}}.  Using the result in (\ref{mixtureALD}), where we denote $\bs \beta_{\tau}\bs X_i=\bs \mu$ without loss of generality,  for each component  $Y_j$  the following holds 

\ber \label{exp1}
Y_j&=& \mu_j + \delta_j\xi_j W + \delta_j  \sqrt{ W} \sigma_j Z_j, \qquad \forall j=1,2..,p 
\eer

\noindent where $Z_j \sim {\cal N}(0,1)$ represents the $j$-th component of $Z$. Then, defining $V=\delta_jW \sim \mbox{Exp}\left( \frac{1}{\delta} \right)$, the representation in (\ref{exp1}) can be also written as

\ber \label{expsigma}
Y_j&=& \mu_j + \xi_j V + \sigma_j \sqrt{\delta_j V} Z_j.
\eer

\noindent Following Kotz et al. (2001) and Kozumi and Kobayashi (2011), and imposing (\ref{constraints}), the result follows since (\ref{expsigma}) represents a univariate AL distribution with location, skewness and scale parameter equal to $\mu_j$, $\tau_j$ and  $\delta_j$, respectively. 

 \hspace{6in} $\square$

\bigskip
\bigskip
%
%

\noindent {\textbf{Proof of Proposition \ref{identifiability}}}
{Remember that $\bs D = \diag [\delta_1, \delta_2,..., \delta_j]$ is an unknown matrix with each $\delta_j >0$, $j=1,...,p$. Moreover $\bs{\tilde \Sigma}= \bs{\tilde \Lambda \Psi \tilde \Lambda} $, where $\bs {\tilde \Lambda}$ is a known $p \times p$ diagonal matrix with $(j,j)$-th element equal to $\tilde \sigma_j= \sqrt{\frac{2}{\tau_j (1- \tau_j)}}$ and where $\Psi$ is an unknown correlation matrix to be estimated. }

\noindent {Suppose we have two sets of parameters (say, $\bs D$ and $\bs {D^*}$, and  $\bs \Psi$ and $\bs{\Psi^*}$), such that 
 \ber \label{ident}\bs{D \tilde \Lambda \Psi \tilde \Lambda D}=\bs{D^* \tilde \Lambda \Psi^* \tilde \Lambda D^*}.\eer
 Then, identifiability analysis asks whether it is  possible to  identify both $\bs D$ and $\Psi$ uniquely.  That is, if (\ref{ident}) holds for some $\bs D \ne \bs D^*$  and some $\bs{\Psi} \ne \bs{\Psi^*}$, then the model is not identifiable, as two different sets of parameters give the same MAL distribution. Now, notice that (\ref{ident}) implies that 
\ber \label{identDiag}
\diag \left( \bs{D \tilde \Lambda \Psi \tilde \Lambda D}\right) = \diag \left( \bs{D^* \tilde \Lambda \Psi^* \tilde \Lambda D^*} \right).
\eer
Since both $\bs \Psi$ and $\bs{\Psi^*}$ are correlation matrices then,  for each $j$-th diagonal element we have $\Psi_{jj}= \Psi_{jj}^*=1$, $\forall j=1,...,p$. This implies that (\ref{identDiag}) can be rewritten as 
\ber \label{diag2}
(\delta_j)^2 \sigma^2_j= (\delta_j^*)^2 \sigma^2_j, \qquad j=1,...,p.
\eer
But then, since by assumption each $\delta_j$ (and $\delta_j^*$) are greater than zero, the relationship in (\ref{diag2}) is satisfied if and only if $\delta_j= \delta_j^*$, $\forall j=1,...,p$, or equivalently $\bs D = \bs{D}^*$. Given this, the relationship in (\ref{ident}) reduces to 
\berr
\bs{D \tilde \Lambda \Psi \tilde \Lambda D}=\bs{D \tilde \Lambda \Psi^* \tilde \Lambda D}.
\eerr
Finally, since both $\bs D$ and $\tilde \Lambda$ are squared diagonal matrices (hence invertible), we have
\berr
( \bs{D \tilde \Lambda} )^{-1}\bs{D \tilde \Lambda \Psi \tilde \Lambda D}( \bs{D \tilde \Lambda} )^{-1}=( \bs{D \tilde \Lambda} )^{-1}\bs{D \tilde \Lambda \Psi^* \tilde \Lambda D}( \bs{D \tilde \Lambda} )^{-1},
\eerr
which implies that $\bs \Psi = \bs {\Psi^*}$. Therefore, both $\bs D$ and $\bs \Psi$ are just identified, as (\ref{ident}) holds if and only if  $\bs D = \bs{D}^*$ and $\bs \Psi = \bs {\Psi^*}$.
}

 \hspace{6in} $\square$
\bigskip

\bigskip

\noindent \textbf{Proof of Proposition \ref{expectedloglik}}. 
 Notice that, under the constraints in (\ref{constraints}),  the representation in (\ref{mixtureALD}) implies that 

\ber
\bs Y_i|W_i=w_i &\sim & {\cal N}_p(\bs \beta_{\tau}\bs X_i + \bs D\tilde{\bs \xi} w_i \, , \,w_i \bs D \tilde{ \bs \Sigma} \bs D ) \label{hier1}
\\
W_i&\sim& Exp(1) \label{hier2}
\eer

\noindent This implies that the  joint density function of $\bs Y$ and $W$ is 

\ber
\label{joint}
f_{Y,W}(\bs y,w)= \frac{\exp{\left\{ (\bs y - \bs \mu) ' \bs D^{-1}\bs \Sigma^{-1} \bs \xi \right\}}}{(2 \pi)^{p/2} |\bs S|^{1/2}} \left(  w^{-p/2} \exp{\left\{ -\frac{1}{2}   \frac{m}{w} - \frac{1}{2} w (d +2)    \right\}}\right).
\eer

\noindent Then, using  (\ref{joint}), the complete log-likelihood function (up to additive constant terms) can be written as follows:

\ber \label{completeLik}
l_c(\bs\beta_{\tau}, \bs D_{(\bs \delta)},\tilde {\bs\Sigma}_{(\bs \Psi)}|\bs Y_i,W_i) &=& \sum_{i=1}^n(\bs Y_i-\bs \beta_{\tau}\bs X_i)'\bs D^{-1}_{(\bs \delta)}\tilde{\bs \Sigma}^{-1}_{(\bs \Psi)}\tilde {\bs \xi} - \frac{n}{2}\log |\bs D_{(\bs \delta)} \tilde {\bs{\Sigma}}_{(\bs \Psi)} \bs D_{(\bs \delta)}|  \\
&-&   \frac{1}{2} \sum_{i=1}^n \frac{1}{W_i}(\bs Y_i-\bs \beta_{\tau}\bs X_i)' (\bs D_{(\bs \delta)} \tilde{ \bs \Sigma}_{(\bs \Psi)} \bs D_{\bs \delta})^{-1} (\bs Y_i- \bs \beta_{\tau}\bs X_i) \\
&-&\frac{1}{2}  \tilde{ \bs \xi}' \tilde{ \bs \Sigma}_{(\bs \Psi)} \tilde { \bs \xi} \sum_{i=1}^n W_i .
\eer

\noindent where we use the notation $\bs D_{(\bs \delta)}$ and $\tilde {\bs\Sigma}_{(\bs \Psi)}$ to express the matrices $\bs D$ and $\tilde{\bs \Sigma}$ as a function of their parameters, $\bs \delta$ and $\bs \Psi$, respectively. In practice, $W_i$ is a latent variable and, hence, not observable. For this reason, the E-step of the EM algorithm considers the conditional expectation of the complete log-likelihood function, given the observed data $\bs Y_i$ and the  parameter estimates ($\hat{\bs \beta}_{\bs \tau}, {\bs D}_{(\hat{\bs \delta})}, {\tilde {\bs \Sigma}}_{(\hat{\bs \Psi})}$). That is, 

\ber
E\left[l_c(\bs \beta_{\tau}, \bs D_{(\bs \delta)},\tilde {\bs\Sigma}_{(\bs \Psi)}|\bs Y_i, \hat{\bs \beta}_{\tau}, \bs{D}_{(\hat{\bs \delta})}, {\tilde {\bs\Sigma}}_{(\hat{\bs \Psi})})  \right] &=&  \sum_{i=1}^n(\bs Y_i-\bs \beta_{\tau}\bs X_i)'\bs D^{-1}_{\bs \delta}\tilde{\bs{\Sigma}}^{-1}_{(\bs \Psi)}\tilde{\bs \xi} - \frac{n}{2}\log |\bs D_{(\bs \delta)} \tilde {\bs \Sigma}_{(\bs \Psi)} \bs D_{(\bs \delta)}|\\
&-&  \frac{1}{2} \sum_{i=1}^n E[W_i^{-1}](\bs Y_i-\bs \beta_{\tau}\bs X_i)' (\bs D_{(\bs \delta)} \tilde{ \bs \Sigma}_{(\bs \Psi)} \bs D_{\bs \delta})^{-1} (\bs Y_i- \bs \beta_{\tau}\bs X_i)\nonumber \\[-0.1in]
&&\\
& -&\frac{1}{2}  \tilde{ \bs \xi}' \tilde{ \bs \Sigma}_{(\bs \Psi)} \tilde { \bs \xi}\sum_{i=1}^n E[ W_i] 
\eer

\noindent where $E[W_i| \cdot]$ and $E[W_i^{-1}| \cdot]$ denote the conditional expectations of $W_i$ and $W_i^{-1}$ conditional on $(\cdot)$, respectively.

\noindent  Let $\tilde {m_i}= (\bs y_i- \bs \beta_{\tau}\bs X_i)' (\bs D_{(\bs \delta)} \tilde{ \bs \Sigma}_{(\bs \Psi)} \bs D_{(\bs \delta)})^{-1}(\bs y_i- \bs \beta_{\tau}\bs X_i)$, for every $i=1,2,...,n$, and define  $\tilde d= \tilde {\bs \xi}' \tilde{ \bs \Sigma}_{(\bs \Psi)} \tilde{ \bs \xi}$. Using the joint distribution of $\bs Y$ and $W$ derived in (\ref{joint}) and the pdf of $\bs Y$ given in (\ref{MALdensityCons}), we have that

\ber
f_{W|Y}(W_i|\bs Y_i=\bs y_i) &=& \frac{f_{W,Y}(w_i,\bs y_i)}{f_Y(\bs y_i)} = \frac{w_i^{-p/2} \left(\frac{2+\tilde d}{\tilde m_i} \right)^{\nu/2} \exp{\left\{  -\frac{\tilde m_i}{2w_i}- \frac{w_i(2+\tilde d)}{2}   \right\}}}{2 K_{\nu}\left( \sqrt{(2+\tilde d)\tilde m_i} \right)}
\eer

\noindent which corresponds to a Generalized Inverse Gaussian (GIG) distribution with parameters $\nu, {2+\tilde d}, \tilde{m_i}$, i.e.\footnote{The pdf of a GIG($p,a,b$) distribution is defined as $f_{GIG}(x; p,a,b)= \frac{\left(\frac{a}{b}\right)^{p/2}}{2K_{p}(\sqrt{ab})} x^{p-1} e^{-\frac{1}{2}\left( ax +bx^{-1}  \right)}$, with $a>0$, $b>0$ and $p \in {\cal R}$.}

\ber
f_{W|Y}(W_i|\bs Y_i=\bs y_i) \sim \mbox{GIG}\left(\nu, \tilde d+2,  \tilde{m_i}\right).
\eer

\noindent It follows that 

\ber \label{wi}
 E [W_i|\bs Y_i, \hat{\bs \beta}_{\tau}, \bs{D}_{(\hat{\bs \delta})}, {\tilde {\bs\Sigma}}_{(\hat{\bs \Psi})}) ] = \left(  \frac{\hat {\tilde {m}}_i}{2+\hat {\tilde d}}  \right)^{\frac{1}{2}} \frac{K_{\nu +1}\left( \sqrt{(2+\hat{\tilde d})\hat {\tilde {m}}_i} \right)}{K_{\nu}\left(   \sqrt{(2+\hat {\tilde d}) \hat {\tilde {m}}_i}\right)}
\eer

\noindent and 

\ber \label{wi-1}
E [W_i^{-1}|\bs Y_i, \hat{\bs \beta}_{\tau}, \bs{D}_{(\hat{\bs \delta})}, {\tilde {\bs\Sigma}}_{(\hat{\bs \Psi})}) ] = \left( \frac{2+\hat {\tilde d}}{\hat {\tilde {m}}_i} \right)^{\frac{1}{2}}  \frac{K_{\nu +1}  \left( \sqrt{(2+\hat {\tilde d})\hat {\tilde {m}}_i} \right)}{K_{\nu}  \left( \sqrt{(2+\hat {\tilde d})\hat {\tilde {m}}_i} \right)} - \frac{2 \nu}{\hat {\tilde {m}}_i}  
\eer

\noindent where

\ber
\hat {\tilde{m}}_i &=& (\bs y_i- \hat{\bs \beta}_{\tau}\bs X_i)' (\bs D_{(\hat{\bs \delta})} {\tilde {\bs \Sigma}}_{\hat{\bs \Psi}} \bs D_{(\hat{\bs \delta})})^{-1}(\bs y_i- \hat {\bs \beta}_{\tau}\bs X_i) \\
\hat{ \tilde d} &=& \tilde {\bs \xi}' {\tilde {\bs \Sigma} }_{(\hat{\bs \Psi})}\tilde{ \bs \xi}.
\eer

\noindent Denoting  the two conditional expectations in (\ref{wi}) and (\ref{wi-1}) by $u_i$ and $z_i$ respectively, conclude the proof. 
\hspace{6in} $\square$

\bigskip

\bigskip

\noindent \textbf{Proof of Proposition \ref{Mestimates}}. 
\noindent Imposing the first order conditions on (\ref{OF1}) - (\ref{OF3}) with respect to $\bs {\beta}_{\tau}$ and $\tilde{\bs  \Sigma}_{(\bs \Psi)}$ gives the parameter estimates in (\ref{focBeta}), (\ref{focSigma}) and (\ref{focD}).

\end{document}